\documentstyle[prd,aps,epsf,floats]{revtex} 
\draft

\begin{document}
\twocolumn[\hsize\textwidth\columnwidth\hsize\csname
@twocolumnfalse\endcsname
\title{\vbox{\normalsize\rm\rightskip=0cm\leftskip=0cm\noindent 
     Minor update of my contribution to the 
     Proceedings of the Summer School on Physics with
     Neutrinos (Zuoz, Switzerland, August 4--10, 1996), edited by 
     M.~P.~Locher (Paul Scherrer Institut, Villingen, 1996).\hfil{\ }}
\vskip0.5cm 
Neutrino Masses in Astrophysics and Cosmology}
\author{Georg G.~Raffelt}
\address{Max-Planck-Institut f\"ur Physik,
F\"ohringer Ring 6, 80805 M\"unchen, Germany}
\date{\today}
\maketitle
\begin{abstract}
Astrophysical and cosmological arguments and observations give us the
most restrictive constraints on neutrino masses, electromagnetic
couplings, and other properties. Conversely, massive neutrinos would
contribute to the cosmic dark-matter density and would 
play an important role for the formation of structure in the
universe. Neutrino oscillations may well solve the solar neutrino
problem, and can have a significant impact on supernova physics.
\end{abstract}
\vskip2.1pc]


\section{Introduction}

Within the standard model of elementary particle physics, neutrinos
play a special role in that they are the only fermions that appear
with only two degrees of freedom per family, which are massless, and
which interact only by the weak force apart from gravitation. If
neutrinos had masses or anomalous electromagnetic interactions, or if
right-handed (sterile) neutrinos existed, this would constitute the
long-sought ``physics beyond the standard model.'' Hence the
enthusiasm with which experimentalists search for neutrino
oscillations, neutrinoless double-beta decay, a signature for a
neutrino mass in the tritium beta decay spectrum, or for neutrino
electromagnetic dipole or transition moments.

Over the years, many speculations about hypothetical neutrino
properties and their consequences in astrophysics and cosmology have
come and gone. I shall not pursue the more exotic of those conjectures
such as strong neutrino-neutrino interactions by majoron and other
couplings, small neutrino electric charges, the existence of low-mass
right-handed partners to the established sequential flavors, and so
forth. Any of them can be significantly constrained by astrophysical
and cosmological methods \cite{RaffeltBook}, but currently there does
not seem to be a realistic way to positively establish physics beyond
the standard model on such grounds. Therefore, I will focus on the
more conservative modifications of the standard-model neutrino sector,
namely on neutrino masses and mixings. Surely, the discovery of a
nonvanishing mass is the holy grail of neutrino physics, and one that
actually may be established to exist in the near future.

Arguably the most important astrophysical information about neutrino
properties is the cosmological mass limit of about $40\,\rm eV$ which,
in the case of $\nu_\tau$, improves the direct experimental limit by
about six orders of magnitude. If neutrinos decay, this limit can be
circumvented. However, the only standard-model decay that would be
fast enough is $\nu_\tau\to e^+e^-\nu_e$ if $m_{\nu_\tau}\agt
2m_e$. This mode can be constrained by the absence of $\gamma$ rays
from the supernova (SN) 1987A and other arguments to be far too slow
than needed to evade the cosmological mass limit. Therefore, its
violation requires fast invisible neutrino decays, i.e.\ rather exotic
physics beyond the standard model. These issues are explored in
Sec.~II.

Currently favored models for the formation of structure in the
universe exclude neutrinos as a dark-matter candidate. Still,
neutrinos with a mass of a few eV could play an important positive
role in mixed hot plus cold dark matter scenarios. The chances for a
signature of such scenarios in future cosmic microwave background maps
has been assessed (Sec.~III).

Big-bang nucleosynthesis (BBN) has long been used to constrain the
number of light neutrino species \cite{KolbTurner}, which however is
now well established to be 3 from precision measurements of the $Z^0$
decay width.  More recently, the BBN argument has been revived to
constrain a $\nu_\tau$ mass in the MeV range
\cite{MeVneutrinos}. However, the assumption of a neutrino mass in
excess of the cosmological limit of about $40\,\rm eV$ is somewhat
forced because it requires exotic neutrino interactions
beyond the standard model.

The existence of three massless or nearly massless two-component
neutrino flavors is compatible with standard BBN, even though there is
some current debate about the interpretation of certain observations
which imply somewhat incompatible or inconsistent primordial light
element abundances \cite{BBN}.  However, what is at stake is not so
much any serious implication for neutrino physics, but rather the
precise value of the baryon content $\eta$ of the universe.  In any
event, BBN is not sensitive to those nonstandard neutrino properties
which are most likely to be found in nature, i.e.\ small masses and
mixings.  Therefore, I do not want to embark here on any further
discussion of BBN.

A positive identification of neutrino masses most likely will come
from the discovery of neutrino oscillations. Current indications for
this phenomenon include the solar neutrino problem, the atmospheric
neutrino anomaly, and the LSND excess counts of $\overline\nu_e$'s. As
these issues are discussed by other speakers at this School, I will
give only the briefest of summaries at the beginning of Sec.~IV. For
the most part, that section will be dedicated to the impact of
neutrino oscillations on SN~physics. 


\section{Mass Limits}

\subsection{Cosmological Mass Limit}

\begin{figure}[b]
\centering\leavevmode
\epsfxsize=7cm
\epsfbox{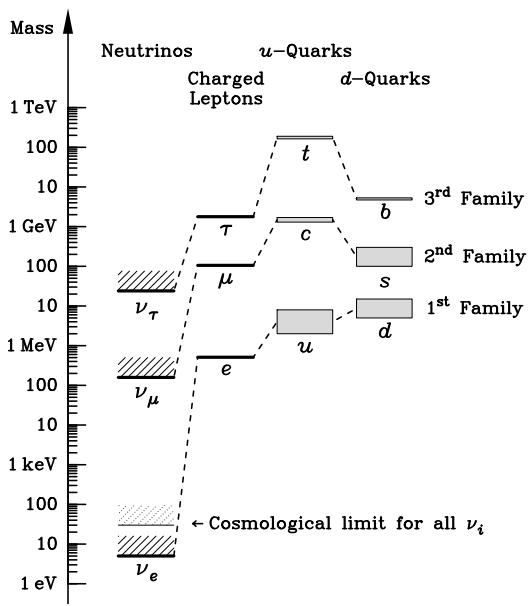}
\medskip
\caption{Mass spectrum of elementary fermions. For neutrinos, the
direct experimental mass limits as well as the cosmological limit are
shown.\label{fig:fermionmasses}}
\end{figure}

The most important cosmological contribution to neutrino physics is
the mass limit which arises from the requirement that the universe is
not ``overclosed'' by neutrinos \cite{KolbTurner}. In the framework of
the big-bang scenario of the early universe one expects about as many
cosmic ``black-body neutrinos'' as there are microwave photons. In
detail, the cosmic energy density in massive neutrinos is found to be
$\rho_\nu=\frac{3}{11}\,n_\gamma\,\sum m_\nu$ with $n_\gamma$ the
present-day density in microwave background photons. The sum extends
over the masses of all sequential neutrino flavors. In units of the
critical density this is
\begin{equation}
\Omega_\nu h^2=\sum \frac{m_\nu}{93\,\rm eV},
\end{equation}
where $h$ is the present-day
Hubble expansion parameter in units of $100\,\rm
km\,s^{-1}\,Mpc^{-1}$. The observed age of the universe together with
the measured expansion rate yields $\Omega h^2\alt 0.4$ so that for
any of the known families
\begin{equation}
m_\nu\alt 40\,{\rm eV}.
\end{equation}
If one of the neutrinos had a mass near this bound it would be the
main component of the long-sought dark matter of the universe.

The importance of this result is illustrated in
Fig.~\ref{fig:fermionmasses} which shows the mass spectrum of the
quarks and charged leptons, and the direct experimental neutrino mass
limits. Except for $\nu_e$, the cosmological limit is far below the
experimental ones which implies that neutrino masses (except
for $\nu_e$) cannot be detected by direct experimental methods. It
also implies that if neutrinos have masses at all, they are very much
smaller than those of the other fundamental fermions. Therefore,
neutrinos play a very special role, even if they were to carry
nonvanishing masses after all, a hypothesis which is entertained by a
majority of all particle physicists.

\subsection{Decaying Neutrinos}

\begin{figure}[b]
\centering\leavevmode
\epsfxsize=8cm
\epsfbox{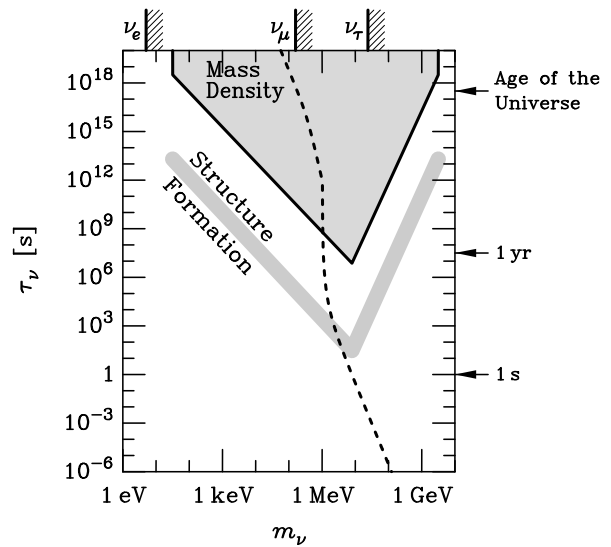}
\medskip
\caption{Cosmological bounds on neutrino masses and lifetimes. The
experimental mass limits are shown above the main panel. The dashed
line is the lifetime for $\nu_\tau\to\nu_e\gamma$ and
$\nu_\tau\to\nu_e e^+ e^-$ under the assumption of maximum 
$\nu_e$-$\nu_\tau$-mixing.
\label{fig:masslifetime}}
\end{figure}

The cosmological mass limit is based on the assumption that neutrinos
are stable which most likely they are not if they have
masses. Sufficiently early decays into nearly massless daughter
particles would allow the energy stored in the massive neutrinos to be
redshifted enough so that the universe would not be ``overclosed''
after all. In Fig.~\ref{fig:masslifetime} the range of neutrino masses
and lifetimes that remains forbidden is shown by the shaded area
marked ``Mass Density.'' A detailed construction of this plot is found
in Ref.~\cite{DicusKolbTeplitz}.

A ``decaying neutrino cosmology'' actually has some attractive
features for the formation of structure in the cosmic matter
distribution. In Sec.~\ref{sec:StructureFormation} below I will
discuss that the standard cold dark matter cosmology has the problem
of producing too much power in the density-fluctuation spectrum on
small scales, and that a mixed hot plus cold dark-matter scenario is
one way of fixing this problem. Another way is with decaying neutrinos
because the universe would become matter dominated when the massive
neutrino becomes nonrelativistic, would return to radiation domination
when it decays, and would become matter dominated again at a later
time. As structure grows by gravitational instability only in phases
of matter domination, one has two more parameters at hand (the
neutrino mass and lifetime) to tune the final density fluctuation
spectrum. For neutrino parameters in the shaded band marked
``Structure Formation'' in Fig.~\ref{fig:masslifetime} this mechanism
could at least partially solve the problems of the cold dark matter
cosmology~\cite{DecayingNeutrinos}.

The snag with this sort of scenario is that within the
particle-physics standard model even massive neutrinos cannot decay
fast enough. Even mixed, massive neutrinos cannot decay at tree-level
by processes of the sort $\nu_\tau\to\nu_e\overline\nu_e\nu_e$ because
of the absence of flavor-violating neutral currents. Therefore, only
radiative decays of the sort $\nu_\tau\to\nu_e\gamma$ are possible.
Because they proceed through a one-loop amplitude, and because of
their so-called GIM suppression, their rate is exceedingly slow.  Even
if the radiative mode were enhanced by some unknown mechanism,
radiative decays can be excluded in a large range of masses and
lifetimes because the final-state photons would appear as
contributions to the cosmic photon backgrounds \cite{KolbTurner}.
Therefore, decaying-neutrino cosmologies as well as a circumvention of
the cosmological mass bound require ``fast invisible decays,'' i.e.\
fast decays with final-state neutrinos or with new exotic particles
such as majorons. Put another way, if neutrinos were found to violate
the cosmological mass bound of $40\,\rm eV$, this would be a signature
for physics ``far beyond'' the standard model. It would require novel
ingredients other than neutrino masses and mixings.

There is one exception to this reasoning if $\nu_\tau$ had a mass
exceeding $2m_e$ because then the decay $\nu_\tau\to\nu_e e^+e^-$ is
kinematically possible. Assuming maximum mixing between $\nu_\tau$ and
$\nu_e$, the lifetime of $\nu_\tau$ as a function of its mass is
plotted in Fig.~\ref{fig:masslifetime} as a dashed line.  (For
$m_{\nu_\tau}<2m_e$ the rate is dominated by
$\nu_\tau\to\nu_e\gamma$.) Evidently, even without exotic extensions
of the standard model, a heavy $\nu_\tau$ could escape the
cosmological mass limit.

\begin{figure}
\centering\leavevmode
\epsfxsize=7cm
\epsfbox{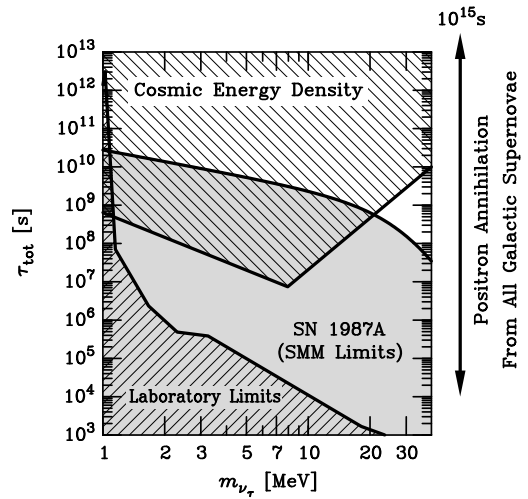}
\medskip
\vskip4pt
\caption{Exclusion areas for the $\nu_\tau$ mass and lifetime if
$\nu_\tau\to\nu_e e^+e^-$ is the only available decay channel. The
``Cosmic Energy Density'' region is taken from
Fig.~\ref{fig:masslifetime}.\label{fig:heavytau}}
\end{figure}

This loophole can be plugged by a combination of laboratory and
astrophysical arguments. First, there are numerous laboratory limits
on the $\nu_\tau$-$\nu_e$ mixing angle. In Fig.~\ref{fig:heavytau} I
show exclusion areas under the assumption that $\nu_\tau\to\nu_e e^+
e^-$ is the only available decay channel which is open due to
$\nu_\tau$-$\nu_e$ mixing.  The hatched area marked ``Cosmic Energy
Density'' is the corresponding exclusion area taken from
Fig.~\ref{fig:masslifetime}.

It is thought that in a SN collapse the gravitational binding energy
of the newborn neutron star is emitted almost entirely in the form of
neutrinos, and that this energy of about $3{\times}10^{53}\,\rm ergs$
is roughly equipartitioned between the (anti)neutrinos of all flavors.
Even if $m_{\nu_\tau}$ were as large as allowed by laboratory
experiments (about $24\,\rm MeV$) the $\nu_\tau$ emission process
would not be significantly suppressed by threshold effects. Therefore,
supernovae are powerful $\nu_\tau$ sources. The positrons from the
subsequent $\nu_\tau\to\nu_e e^+e^-$ decay would be trapped in the
galactic magnetic field and would have a lifetime against annihilation
of about $10^5\,\rm yr$. Therefore, the galactic positron flux
integrated over all supernovae over such a period would far exceed the
observed value unless the $\nu_\tau$'s either decay very fast (very
close to the SN), or else they live so long that they escape from the
galaxy before decaying. The excluded range of lifetimes according to
Ref.~\cite{DarI} is indicated in Fig.~\ref{fig:heavytau} by a vertical
arrow.
\looseness=1

A particulary important exclusion range arises from SN~1987A which is
the first and only supernova from which neutrinos were observed. The
gamma-ray spectrometer (GRS) on the solar maximum mission (SMM)
satellite which was operational at the time did not observe any excess
photon counts in coincidence with the neutrino signal, allowing one to
derive restrictive limits on neutrino radiative decays
\cite{RaffeltBook}. For the present discussion it is most interesting
that the absence of observed $\gamma$-rays also allows one to restrict
the inner bremsstrahlung process $\nu_\tau\to\nu_e e^+ e^-\gamma$ and
thus the $\nu_\tau\to\nu_e e^+ e^-$ channel 
\cite{DarII}; the excluded parameter range is shaded 
in~Fig.~\ref{fig:heavytau}. 
\looseness=1

Interestingly, this SN~1987A exclusion range can be extended by new
$\gamma$-ray observations. The time-of-flight delay of MeV-mass
neutrinos from SN~1987A is so large that one could still observe
$\gamma$-rays today; one does not depend on the SMM observations which
were coincident with the neutrino signal. The COMPTEL $\gamma$-ray
telescope has been used for that purpose with two dedicated viewing
periods in 1991 with a total observation time of
$6.82{\times}10^5\,\rm sec$~\cite{Miller}. Thus far, only an analysis
for the $\nu_\tau\to\nu_e\gamma$ channel has been presented. However,
for MeV-mass $\nu_\tau$'s one would expect a dramatic improvement of
the limits on the $\nu_\tau\to\nu_e e^+ e^-\gamma$ channel as well;
such an analysis is in progress~\cite{MillerII}.

In summary, if one extends the particle-physics standard model only
with neutrino masses and mixings, the cosmological mass bound remains
firm as there is no viable neutrino decay channel which is both fast
enough and ``invisible.'' Conversely, if neutrino masses in excess of
about $40\,\rm eV$ were to show up in experiments, this would indicate
novel neutrino interactions far outside of what is expected by the
standard model. In this case decaying neutrinos could also play a
useful role for the formation of structure in the universe.

\subsection{Supernova Mass Limits}

For the sake of completeness, two mass limits deserve mention which
were derived from the SN~1987A neutrino signal. First, the absence of
a discernible time-of-flight dispersion of the observed
$\overline\nu_e$ burst gave rise to $m_{\nu_e}\alt20\,\rm eV$
\cite{SNmasslimit}. This limit is now obsolete in view of the improved
experiments concerning the tritium $\beta$ decay endpoint
spectrum. Even though these results seem to be plagued with
unidentified systematic errors, a $\nu_e$ mass as large as $20\,\rm
eV$ does not seem to be viable.
 
Second, if neutrinos had Dirac masses, helicity-flipping collisions in
the dense inner core of a SN would produce right-handed states.
Because these sterile neutrinos are not trapped they carry away the
energy directly which otherwise escapes by a diffusion process to the
neutrino sphere from where it is radiated by standard left-handed
neutrinos. Therefore, the energy available for standard neutrino
cooling would be diminished, leading to a shortening of the SN~1987A
$\overline\nu_e$ burst. Because the burst duration roughly agrees with
theoretical expectations, this scenario can be constrained, leading to
$m_\nu\alt 30\,\rm keV$ on a possible Dirac mass for the $\nu_\mu$
and $\nu_\tau$ \cite{RaffeltBook}. Of course, such a large mass would
violate the cosmological limit and thus is only of interest if there
are fast invisible decays beyond the standard model.  Typically, even
``invisible'' decay channels would involve (left-handed) final-state
neutrinos which could become visible in the detectors which registered
the SN~1987A signal.  Because the sterile neutrinos which escape
directly from the SN core would have higher energies than those
emitted from the neutrino sphere, these events would stick out from
the observed SN~1987A signal.  This allows one to derive additional
limits on certain decay channels of Dirac-mass $\nu_\mu$'s and
$\nu_\tau$'s \cite{Dodelson}.

Of course, much improved mass limits could be derived if one were to
observe a future galactic supernova. In a detector like the proposed
Supernova Burst Observatory (SNBO) one could observe $\nu_\mu$'s and
$\nu_\tau$'s by a coherently enhanced neutral-current nuclear
dissociation reaction of the type $\nu+(Z,N)\to (Z,N-1)+n+\nu$. In
principle, one could be sensitive to time-of-flight signal dispersion
effects corresponding to neutrino masses of a few $10\,\rm eV$ for
$\nu_\mu$ or $\nu_\tau$, especially if the SNBO neutral-current signal
were analysed in conjunction with the charged-current
$\overline\nu_ep\to ne^+$ signal expected for the Superkamiokande
detector \cite{Cline}.


\section{Neutrinos as Dark Matter}

\subsection{Galactic Phase Space}

Cosmology implies a mass limit of about $40\,\rm eV$ on all
sequential neutrinos. If this limit were saturated by one of the
neutrinos, say the $\nu_\tau$, it would constitute the dark matter of
the universe. Is this possible and plausible? The current standard
answer is ``no'' because neutrinos as dark matter candidates fare
poorly on two main grounds. 

The first is a well-known problem with neutrinos filling the
dark-matter haloes of galaxies. By definition, galactic dark-matter
neutrinos would be gravitationally bound to the galaxy so that their
velocity would be bound from above by the galactic escape velocity
$v_{\rm esc}$, yielding an upper limit on their momentum of $p_{\rm
  max}=m_\nu v_{\rm esc}$. Because of the Pauli exclusion principle
the maximum number density of neutrinos is given when they are
completely degenerate with a Fermi momentum $p_{\rm max}$, i.e.\ it is
$n_{\rm max}=p_{\rm max}^3/3\pi^2$. Therefore, the maximum local mass
density in dark-matter neutrinos is $m_\nu n_{\rm max}= m_\nu^4 v_{\rm
  esc}^3/3\pi^2$. As this value must exceed a typical galactic dark
matter density, one obtains a {\it lower\/} limit on the necessary
neutrino mass. A refinement of this simple derivation is known as the
Gunn-Tremaine limit \cite{GunnTremaine}.  For typical spiral galaxies
it is in the range of a few $10\,\rm eV$.

Therefore, dark-matter neutrino masses are squeezed between the upper
limit from the overall cosmic mass density, and the lower limit from
the galactic phase-space argument.  They are squeezed, but perhaps not
entirely squeezed out. Neutrinos could not constitute the dark matter
of dwarf galaxies where a mass of a few $100\,\rm eV$ is required by
the Gunn-Tremaine argument. However, perhaps the dark matter in dwarf
galaxies is of a different physical nature. At any rate, the galactic
phase-space argument surely disturbs any simple-minded fantasy about
neutrinos being the dark matter on all scales.

\subsection{Structure Formation}
\label{sec:StructureFormation}

\begin{figure}[ht]
\centering\leavevmode
\epsfxsize=\hsize
\epsfbox{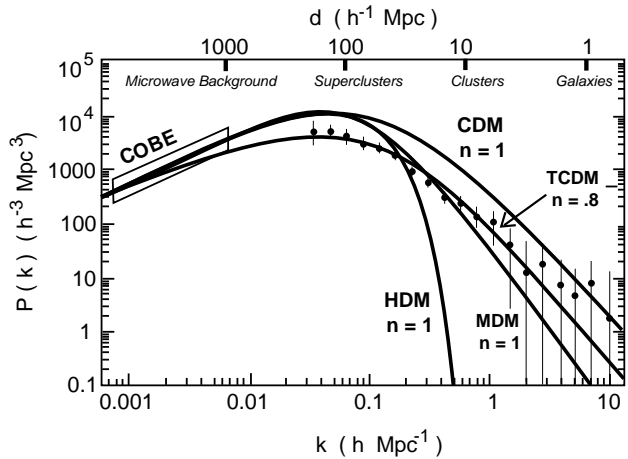}
\medskip
\caption{Comparison of matter-density power spectra for cold dark
matter (CDM), tilted cold dark matter (TCDM), hot dark matter (HDM),
and mixed hot plus cold dark matter (MDM) for large-scale structure
formation \protect\cite{Steinhardt}. All theoretical curves are
normalized to COBE and include only linear approximation; nonlinear
corrections become important on scales smaller than about 
$10\,\rm Mpc$.\label{fig:powerspectrum}}
\end{figure}

The main argument against neutrino dark matter arises from current
scenarios of how structure forms in the cosmic matter distribution.
One pictures a primordial power spectrum of low-amplitude density
fluctuations which are later amplified by the action of gravity. The
expected final distribution of galaxies then depends on both the
nature of the dark matter and the original fluctuation spectrum.  The
result of this sort of reasoning are often displayed in a plot like
Fig.~\ref{fig:powerspectrum} where the Fourier transform of the matter
distribution is shown as a function of wave-number or length scale.
The data are derived from the analysis of observed galaxy
distributions.

Inflationary models of the early universe predict a roughly
scale-invariant primordial fluctuation spectrum
(Harrison-Zeldovich-spectrum). At the time of matter-radiation
decoupling its amplitude is normalized by the COBE observations of
angular temperature variations in the cosmic microwave background.
From Fig.~\ref{fig:powerspectrum} it is evident that a standard cold
dark matter (CDM) scenario if normalized to COBE predicts more power
in the small-scale galaxy distribution than is actually observed.

Neutrinos, on the other hand, represent so-called hot dark matter
because they stay relativistic almost until the epoch of radiation
decoupling. This implies that their relativistic free streaming erases
the primordial fluctuation spectrum on small scales, suppressing the
formation of small-scale structure (Fig.~\ref{fig:powerspectrum}).

Of course, neutrinos as dark matter may still be viable if the
original seeds for structure formation are not provided by
inflation-induced initial density fluctuations, but rather by
something like cosmic strings or textures \cite{Brandenberger}.  Such
scenarios involving topological defects do not seem to be excluded,
but they are certainly disfavored by current main-stream cosmological
thinking, and have not been quantitatively worked out in comparable
detail as the CDM-type cosmologies.

The problem of a standard CDM cosmology depicted in 
Fig.~\ref{fig:powerspectrum} can be patched up in a variety of
ways. One is to tinker with the primordial spectrum of density
fluctuation which may have been almost, but not quite, of the
Harrison-Zeldovich form. One example of such a ``tilted cold dark
matter'' (TCDM) result is shown in Fig.~\ref{fig:powerspectrum}.

Another patch-up is to invoke a mixed hot plus cold dark matter (MDM)
cosmology (Fig.~\ref{fig:powerspectrum}) where the hot component
erases enough of the initial power spectrum on small scales to
compensate for the overproduction of small-scale power of pure CDM.
In an $\Omega=1$ universe, the best fit is obtained with a total mass
in neutrinos corresponding to $\sum m_\nu=5\,\rm eV$ with an
equipartition of the masses among the flavors \cite{HDM}. Primack has
argued that a MDM cosmology not only fixes the problems with the power
spectrum, but also avoids an overdensity of the dark matter in central
galactic haloes~\cite{Primack}.

\subsection{Cosmic Microwave Background}

\begin{figure}[ht]
\vskip-1.3cm
\centering\leavevmode
\epsfxsize=5cm
\epsfbox{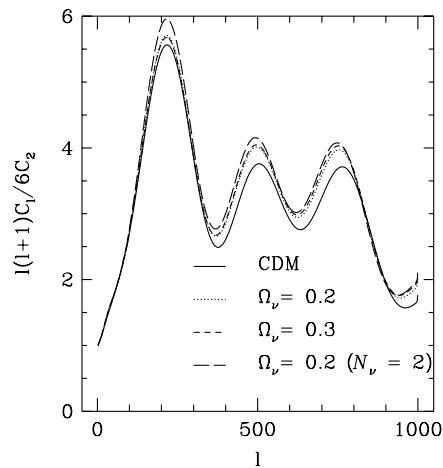}
\vskip0.6cm
\caption{Power spectrum of the temperature sky map for the cosmic
microwave background in a cold dark matter cosmology, and three
variants of mixed dark matter~\protect\cite{Stebbins}. 
\label{fig:cmb}}
\end{figure}

Granted that something like a CDM cosmology describes our universe,
how will we ever know if indeed it contains a small component of
neutrino dark matter? One new source of information will come in the
form of the precision sky maps of the temperature variations of the
cosmic microwave background that will be obtained by the MAP and the
Planck Surveyor (formely COBRAS/SAMBA) 
satellites. Such sky maps are usually interpreted in
terms of their multipole expansion. The expected power as a function
of the multipole order $l$ is shown in Fig.~\ref{fig:cmb} for a pure
cold dark matter cosmology as a solid line according to
Ref.~\cite{Stebbins}. The modified power spectra for three versions of
a mixed hot plus cold dark matter cosmology are also shown.

The first current ambition of cosmic microwave experiments is to
identify the first of the ``Doppler peaks'' in the power spectrum.
With the high angular resolution planned for the Planck Surveyor,
however, one will be able, in principle, to distinguish between the
CDM and the MDM curves shown in Fig.~\ref{fig:cmb}. However, there are
other unknown cosmological parameters such as the overall mass
density, the Hubble constant, the cosmological constant, and so forth,
which all affect the expected power spectrum. All of these parameters
will have to be determined by fitting the power spectrum obtained from
future measurements. Therefore, it remains to be seen if a small
neutrino component of the overall dark matter density can be
identified by cosmic microwave data.


\section{Neutrino Oscillations}

\subsection{Evidence So Far}

\begin{figure}[ht]
\centering\leavevmode
\epsfxsize=\hsize
\epsfbox{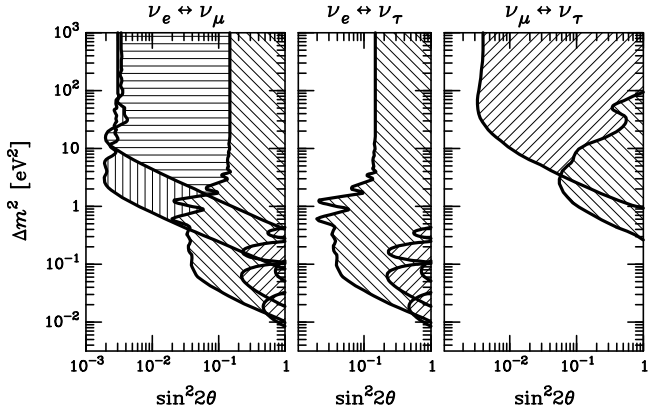}
\medskip
\caption{Limits on neutrino masses and mixing angles from laboratory
experiments. For detailed references see
Refs.~\protect\cite{RaffeltBook,ParticleData}.
\label{fig:oscillations}}
\end{figure}

While neutrino masses would play a very important role in cosmology,
it appears unlikely that cosmological arguments or observations alone
will be able to prove or disprove this hypothesis anytime soon.
Therefore, the only realistic and systematic path to search for
neutrino masses is to search for neutrino oscillations as explained by
other speakers at this school.  Unsurprisingly, a vast amount of
experimental effort is dedicated to this end. While large regions of
neutrino mass differences and mixing angles have been excluded
(Fig.~\ref{fig:oscillations}) there is yet no uncontestable positive
signature for oscillations. However, there exist a number of
experimental ``anomalies'' that could well point to oscillations.

\begin{figure}[ht]
\centering\leavevmode
\epsfxsize=6cm
\epsfbox{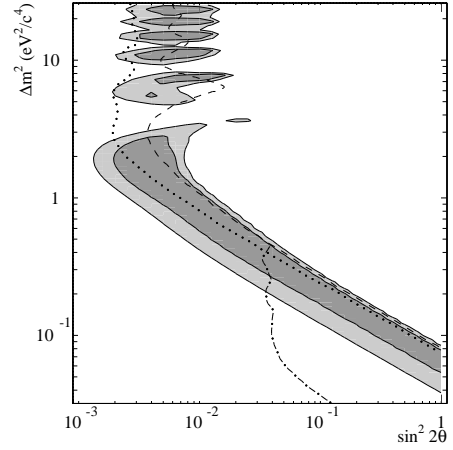}
\medskip
\caption{The LSND favored mixing parameters \protect\cite{LSND}.  The
shaded areas are the 90\% or 99\% likelihood regions. Also shown are
$90\%$ C.L.\ limits from KARMEN (dashed curve), BNL-E776 (dotted
curve), and the Bugey reactor experiment (dot-dashed curve).
\label{fig:LSND}}
\end{figure}

The most recent example is a pure laboratory experiment at Los Alamos
where neutrinos are produced in a proton beam dump. The secondary
positive pions decay according to $\pi^+\to\mu^++\nu_\mu$ and the
muons according to $\mu^+\to e^++\overline\nu_\mu+\nu_e$. In the
Liquid Scintillator Neutrino Detector (LSND) about 30 meters
downstream, a significant number of excess
$\overline\nu_e$ counts was obtained \cite{LSND} which cannot be due
to the primary source which does not produce $\overline\nu_e$'s. These
excess counts can be interpreted as the appearance of oscillated
$\overline\nu_\mu$'s (Fig.~\ref{fig:LSND}). 
If this interpretation were correct, the
$\nu_e$-$\nu_\mu$ mass difference could be of order $1\,\rm eV$ or
more, pointing to cosmologically significant neutrino masses. At
the present time one has to wait and see if more LSND data and other
experiments such as KARMEN will confirm this rather tentative finding.

\begin{figure}[b]
\centering\leavevmode
\epsfxsize=\hsize
\epsfbox{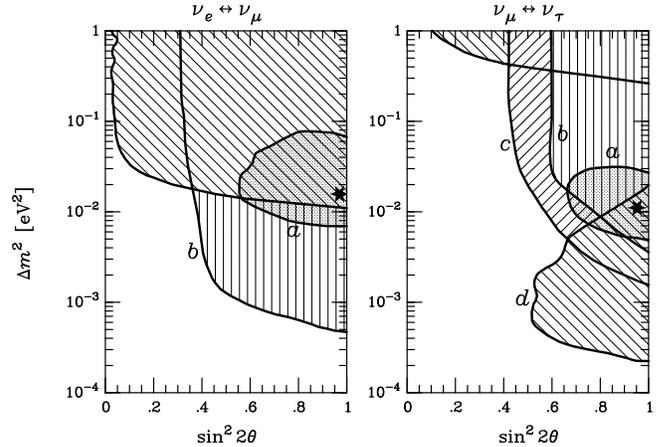}
\medskip
\caption{Limits on neutrino masses and mixing angles from atmospheric
neutrinos. (a) The shaded area is the range of masses and mixing
angles required to explain the $\nu_e/\nu_\mu$ anomaly at Kamiokande
\protect\cite{Fukuda}; the star marks the best-fit values. The hatched
areas are excluded by: (b) $\nu_e/\nu_\mu$ ratio at Fr\'ejus
\protect\cite{Frejus}.  (c) Absolute rate and (d) stopping fraction of
upward going muons at IMB \protect\cite{Becker}. Also shown are the
parameter regions excluded by laboratory
experiments.\label{fig:atmospheric}}
\end{figure}

Another indication for oscillations arises from the atmospheric
neutrino anomaly. The production process by high-energy cosmic-ray
protons is very similar to the LSND experiment, except that the
higher-energy protons produce both positively and negatively charged
pions and kaons in roughly equal proportions so that one expects about
equally many neutrinos as antineutrinos, and a $\nu_\mu:\nu_e$ flavor
ratio of about $2:1$. However, the Kamiokande detector has observed a
flavor ratio more like $1:1$~\cite{Hirata}.  Further, Kamiokande has
seen an angular dependence of the flavor ratio as expected for
oscillations due to the different path lengths through the Earth from
the atmosphere to the detector~\cite{Fukuda}.  In principle, these
observations can be explained by either $\nu_\mu$-$\nu_e$
oscillations, or by $\nu_\mu$-$\nu_\tau$ ones
(Fig.~\ref{fig:atmospheric}).  Either way, nearly maximum mixing is
required with a mass difference of about $10^{-1}\,\rm eV$.  However,
for $\nu_\mu$-$\nu_e$ oscillations the favored range of parameters is
excluded entirely by the nonobservation of a flavor anomaly in
Fr\'ejus \cite{Frejus}, while the $\nu_\mu$-$\nu_\tau$ case is only
partially excluded. Further inconsistencies seem to exist with certain
IMB measurements of the stopping fraction of upward-going
muons~\cite{Becker}. One may hope that the new Superkamiokande
detector will soon clarify this confusing situation.

\begin{figure}
\centering\leavevmode
\epsfxsize=\hsize
\epsfbox{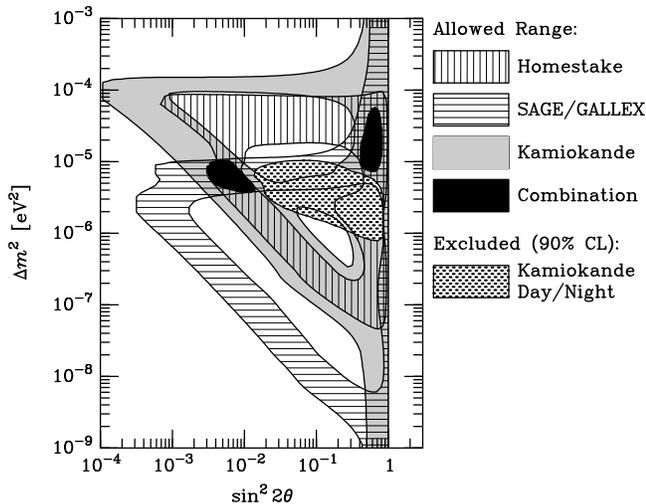}
\medskip
\caption{MSW solutions to the solar neutrino problem according to
Ref.~\protect\cite{HataHaxton} if the flux deficit relative to the
Bahcall and Pinsonneault (1995) solar model is interpreted in terms of
neutrino oscillations.
\label{fig:msw}}
\end{figure}

\begin{figure}
\centering\leavevmode
\epsfxsize=6.8cm
\epsfbox{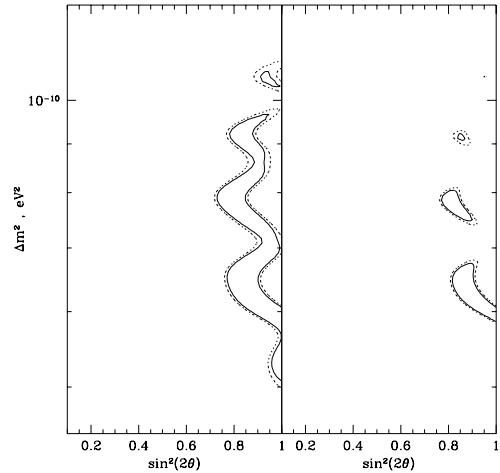}
\medskip
\caption{Vacuum solutions to the solar neutrino problem at 90 and 95\%
C.L.\ according to Ref.~\protect\cite{Krastev} for $\nu_e$-$\nu_\mu$
or $\nu_e$-$\nu_\tau$ oscillations. The two panels represent somewhat
different versions of the statistical analysis.
\label{fig:vacuum}}
\end{figure}

Probably the most convincing tentative evidence for neutrino
oscillations arises from the solar neutrino problem which has been
amply covered by other speakers at this school. If the measured flux
deficits are interpreted in terms of matter-induced oscillations
(MSW-effect) one obtains the favored mixing parameters for
$\nu_e$-$\nu_\mu$ or $\nu_e$-$\nu_\tau$ oscillations as indicated in
Fig.~\ref{fig:msw}. A consistent interpretation in terms of
``long-wavelength'' vacuum oscillations is also possible; the favored
mixing parameters obtained from a typical analysis are shown in
Fig.~\ref{fig:vacuum}. If any of these solutions will indeed bear out
from future solar neutrino experiments remains to be seen. Certainly,
at the present time there is no plausible alternate explanation on the
market.

These indications for neutrino oscillations require three different
mass differences which are not compatible with each other. Therefore,
not all of these results can indeed indicate neutrino oscillations
unless one appeals to the existence of neutrino degrees of freedom
beyond the known sequential flavors, i.e.\ to the existence of sterile
neutrinos. It remains to be seen which (if any) of these results will
withstand closer scrutiny by better data.

Meanwhile it remains of interest to look for other scenarios where
neutrino oscillations could be important. Neutrinos dominate the
dynamics of the early universe and so it is natural to wonder if
oscillations could have significant effects there. However, because
all flavors are in thermal equilibrium with each other, the usual
flavor oscillations would not change anything.  Oscillations into
sterile neutrinos would be a nontrivial effect, but since there is
little theoretical or experimental motivation to speculate about the
existence of low-mass right-handed neutrinos, I will not discuss
neutrino oscillations in the early universe any further.

\subsection{Supernova Physics}

Concentrating on flavor oscillations between sequential neutrinos,
supernovae are natural environments to scrutinize for nontrivial
consequences. A type~II supernova occurs when a massive star ($M\agt
8\,M_\odot$) has reached the end of its life. At this point it
consists of a degenerate iron core, surrounded by several shells of
different nuclear burning phases. Because iron is 
the most tightly bound
nucleus, it cannot gain further energy by nuclear fusion so that no
further burning phase can be ignited at the center. As the iron core
grows in mass because nuclear burning at its surface produces more
``ashes,'' it eventually reaches its Chandrasekhar limit of about
$1.4\,M_\odot$, i.e.\ the maximum mass that can be supported by
electron degeneracy pressure. The subsequent core collapse is halted
only when nuclear densities are reached where the equation of state
stiffens. At this point a shock wave forms at the edge of the inner
core, i.e.\ at the edge of that part of the iron core which collapses
subsonically and thus is in good hydrodynamic ``communication'' with
itself. This shock wave advances outward, and eventually expels the
mantle and envelope of the collapsed object, an event which is
observed as the optical supernova explosion. Essentially, the
gravitational implosion of the core is transformed into an explosion
of the outer parts of the star by the ``shock and bounce'' mechanism.

Virtually all of the binding energy of the newly formed compact star
(about $3{\times}10^{53}\,\rm ergs$) is radiated away by
neutrinos. However, because the collapsed core is so hot and dense
that even neutrinos are trapped, this process takes several seconds
which corresponds to a neutrino diffusion time scale from the center
of the core to the ``neutrino sphere'' at its surface from where these
particles can escape freely. It is thought that the released energy is
roughly equipartitioned between all (anti)neutrino flavors, and that
it is emitted with roughly thermal spectra. 
\looseness=1

In spite of the approximate flavor equipartition of the emitted
energy, neutrino oscillations can have important consequences for
supernova physics because the spectra are different between the
different flavors. Various studies find that the average expected
neutrino energy from a SN is \cite{Janka93}
\begin{equation}\label{eq:energies}
\langle E_\nu\rangle=\cases{10{-}12\,{\rm MeV}&for $\nu_e$,\cr
14{-}17\,{\rm MeV}&for $\overline\nu_e$,\cr
24{-}27\,{\rm MeV}&for $\nu_{\mu,\tau}$ and 
               $\overline\nu_{\mu,\tau}$,\cr}
\end{equation}
i.e.\ typically $\langle E_{\nu_e}\rangle\approx \frac{2}{3}
\langle E_{\overline\nu_e}\rangle$ and 
$\langle E_{\nu}\rangle\approx \frac{5}{3}
\langle E_{\overline\nu_e}\rangle$ for the other flavors.  
The different mean energies are explained by the different main
reactions which trap neutrinos, namely
$\nu_e n \to p e^-$, $\overline\nu_e p\to n e^+$, and
$\nu N\to N\nu$ with $N=n$ or $p$. Because the charged-current
reactions have larger cross sections than the neutral-current ones,
and because there are more neutrons than protons, the $\nu_e$'s have
the hardest time to escape and thus emerge from the 
farthest out and thus coldest layers. Still, the radii of the layers
from which the different flavors escape are not too different so that
the relatively large variation of the spectral temperatures between
the flavors and the equipartition of the emitted energy appears to
contradict the Stefan-Boltzmann law. An explanation for this apparent
paradox is given in Ref.~\cite{Janka95}.

It is conceivable that (resonant) oscillations occur outside of the
neutrino sphere so that the spectra between two flavors can be
swapped. Two possible consequences of such a spectral exchange have
been discussed in the literature.

The first has to do with the explosion mechanism for type~II
supernovae which does not work quite as simple as described
above. Because the shock wave forms at the edge of the subsonic inner
core, not at the edge of the iron core, it moves through a layer of
iron before reaching the stellar mantle. By dissociating iron it
loses energy and stalls after a few $100\,\rm ms$ in typical 
calculations. The deposition of energy by neutrinos which emerge from
the inner core is thought to revive the shock wave so that it
resumes its outward motion and eventually explodes the outer
star. However, this ``delayed explosion mechanism'' still does not
seem to work in typical calculations because the transfer of energy
from the neutrinos to the shock wave is not efficient enough. Recently
the importance of convection both within the neutron star and between
the neutron star and the shock wave has been recognized to play some
role at helping to transfer more energy to the shock wave, but even
then successful explosions are not guaranteed. 

\begin{figure}[ht]
\centering\leavevmode
\epsfxsize=7cm
\epsfbox{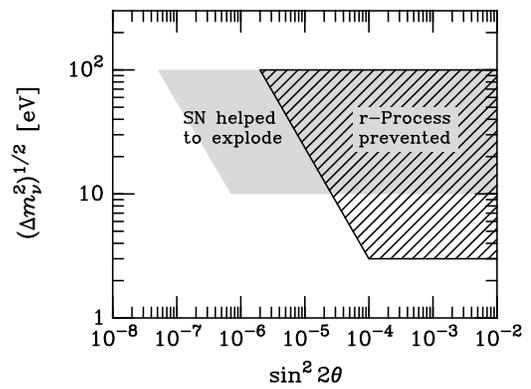}
\medskip
\caption{Mixing parameters between $\nu_e$ and $\nu_\mu$ or $\nu_\tau$
where a spectral swap by resonant oscillations would be efficient
enough to help explode supernovae (schematically after
Ref.~\protect\cite{Fuller92}), and where it would prevent r-process
nucleosynthesis (schematically after Ref.~\protect\cite{Qian}).
\label{fig:snoscillations}}
\end{figure}

If neutrinos follow a ``normal'' mass hierarchy so that $\nu_e$ is
dominated by the lighest mass eigenstate, one can have MSW-type
resonant oscillations between, say, $\nu_e$ and $\nu_\tau$. If this
occurs between the neutrino sphere and the stalling shock wave, the
$\nu_e$'s reaching the shock wave are really oscillated $\nu_\tau$'s
and thus have the higher spectral energies characteristic for that
flavor. The total energy flux in both flavors is about the same, but
the absorption cross sections are larger for larger energies so that
more energy is transferred to the shock wave \cite{Fuller92}. Because
the MSW transition must occur rather close to the neutrino sphere
where the matter densities are large, neutrino mass differences in the
cosmologically interesting regime are required for this effect to
operate. In Fig.~\ref{fig:snoscillations} the approximate range of
masses and mixing angles is shown where neutrino oscillations would
help to explode supernovae.

A second consequence of oscillations is their possible impact on
nucleosynthesis. It has long been thought that the isotopes with
$A\agt70$ are formed by neutron capture which thus requires a
neutron-rich environment. It is now thought that an ideal site for the
r-process is the high-entropy ``hot bubble'' in a SN between the young
neutron star and the advancing shock wave a few seconds after
collapse. The neutrino-driven wind in this dilute environment is
shifted to a neutron-rich phase by $\beta$ processes and because of
the energy hierarchy $\langle E_{\nu_e}\rangle<\langle
E_{\overline\nu_e}\rangle$.  However, if oscillations would cause a
spectral swap between, say, $\nu_e$ and $\nu_\tau$ then this energy
hierarchy would be inverted and the wind would be shifted to a
proton-rich phase, preventing the occurrence of r-process
nucleosynthesis \cite{Qian}. Because this argument applies to a later
phase than the explosion argument above, the neutron star has
thermally settled so that the matter gradients at its surface are much
steeper than before. This makes it harder to meet the adiabaticity
condition for the MSW effect, reducing the range of mixing angles
where this effect operates (Fig.~\ref{fig:snoscillations}).

At the present time it is not certain if r-process nucleosynthesis
indeed occurs in supernovae so that the hatched are in 
Fig.~\ref{fig:snoscillations} cannot be taken as a serious exclusion
plot for neutrino mixing parameters. Still, it is fascinating that
cosmologically interesting neutrino masses would have a nontrivial
impact on SN physics. At any rate, there is a significant range of
small mixing angles where supernovae could be helped to explode by
oscillations, and r-process nucleosynthesis could still proceed
unscathed.  

\subsection{SN~1987A Signal Interpretation}

Neutrinos from a collapsing star were observed for the first and only
time from SN~1987A which occurred in the Large Magellanic Cloud on 23
February 1987. Naturally, the observed signature would be different
if oscillations had occurred between the source and the detectors.
Two observable effects have been discussed in the literature.

The first relates to the so-called neutronization $\nu_e$ burst which
precedes the main cooling phase. It is produced when the shock wave
breaks through the surface of the iron core, allowing the sudden
release of $\nu_e$'s from the reactions $e^-p\to n\nu_e$ from a layer
encompassing perhaps a few $0.1\,M_\odot$ of matter. Most of the core
is deleptonized and thus neutronized only during the relatively slow
subsequent cooling phase. In the water Cherenkov detectors IMB and
Kamiokande which registered the SN~1987A signal, the $\nu_e e\to
e\nu_e$ scattering reaction would have produced forward-peaked
electrons as a signature for this burst, although one would have
expected only a fraction of an event. As the first event in Kamiokande
does point in the forward direction, it has often been interpreted as
being due to the neutronization burst.

\begin{figure}[ht]
\centering\leavevmode
\epsfxsize=6cm
\epsfbox{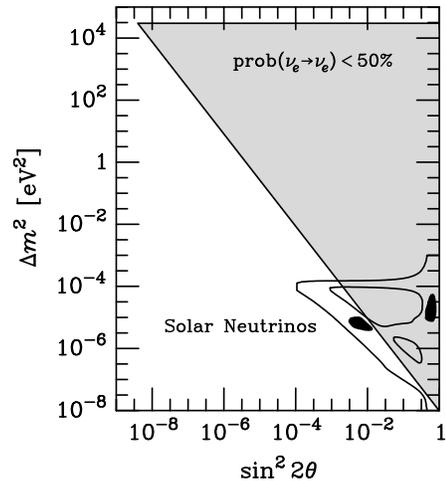}
\medskip
\caption{Mixing parameters between $\nu_e$ and $\nu_\mu$ or $\nu_\tau$
where the prompt SN neutronization burst of $\nu_e$'s would have
resonantly oscillated into another flavor (after
Ref.~\protect\cite{Dirk}). A normal mass hierarchy is assumed where
$\nu_e$ is dominated by the lightest mass eigenstate.  For
orientation, the Kamiokande solar MSW triangle and the MSW solutions
to the solar neutrino problem are also shown.
\label{fig:dirk}}
\end{figure}

If resonant oscillations in the SN mantle and envelope had occurred,
the deleptonization $\nu_e$'s would have arrived as $\nu_\mu$'s or
$\nu_\tau$'s instead which have a much smaller scattering cross
section on electrons, thus reducing the observable signal. In
Fig.~\ref{fig:dirk} the shaded triangle shows the mixing parameters
for which the oscillation probability would have exceeded 50\%,
assuming a normal neutrino mass hierarchy. For orientation, the MSW
triangle for solar neutrinos in the Kamiokande detector and the MSW
solution to the solar neutrino problem are also shown. Evidently, the
small-angle MSW solution would not be in conflict with the
interpretation that the first SN~1987A Kamiokande event was indeed
$\nu_e$ scattering.  Of course, this single event does not allow one
to reach the opposite conclusion that the large-angle MSW solution was
ruled~out.

Most or probably all of the 19 events at Kamiokande and IMB must have
been due to the $\overline\nu_e p\to n e^+$ reaction.  Assuming again
a normal neutrino mass hierarchy, resonant oscillations could not have
swapped the $\overline\nu_e$ spectra with some other flavor; they
could have affected only the $\nu_e$ spectrum.  Therefore, the
observed events represent the initial $\overline\nu_e$ spectrum at the
source unless the mixing angle is large, allowing for significant
non-resonant oscillation effects.  Large mixing angles in the neutrino
sector are motivated by the large-angle MSW and the vacuum solution to
the solar neutrino problem as well as by the oscillation
interpretation of the atmospheric neutrino anomaly.

\begin{figure}[ht]
\centering\leavevmode
\epsfxsize=6cm
\epsfbox{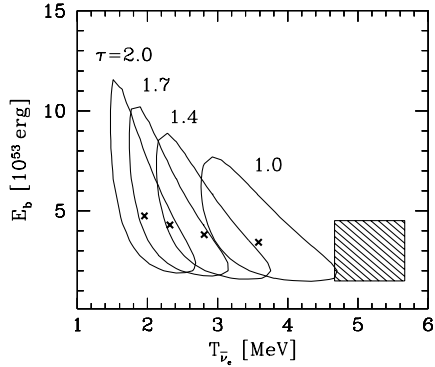}
\medskip
\caption{95\% confidence contours for the neutron star binding energy
and temperature of the primary $\overline\nu_e$ spectrum for the
marked values of $\tau=T_{\overline\nu_\mu}/T_{\overline\nu_e}$ 
\protect\cite{Jegerlehner}. For the neutrino mixing
parameters typical values for the solar vacuum oscillation solution
were chosen ($\Delta m^2=10^{-10}\,\rm eV^2$, $\sin^22\Theta=1$). In
the case $\tau=1$ oscillations do not change the spectra so that this
contour corresponds to the absence of oscillations. The hatched area
represents the range of theoretical predictions.
\label{fig:sn-vacuum}}
\end{figure}

\begin{figure}[ht]
\centering\leavevmode
\epsfxsize=6cm
\epsfbox{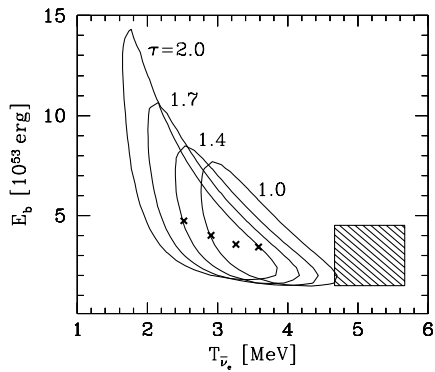}
\medskip
\caption{Same as Fig.~\protect\ref{fig:sn-vacuum} for neutrino mixing
parameters which are typical for the solar large-angle MSW solution
($\Delta m^2=10^{-5}\,\rm eV^2$, $\sin^22\Theta=0.8$). 
\label{fig:sn-msw}}
\end{figure}

One way of interpreting the observed SN~1987A events is to use the
data to derive best-fit values for the total binding energy $E_{\rm
b}$ and the spectral temperature of the observed $\overline\nu_e$'s
which is defined by $T_{\overline\nu_e}=\frac{1}{3}\langle
E_{\overline\nu_e}\rangle$. Assuming certain mixing parameters and
certain relative spectral temperatures
$\tau\equiv T_{\overline\nu_\mu}/T_{\overline\nu_e}$ between the
flavors the results from such an analysis are shown in
Figs.~\ref{fig:sn-vacuum} and \ref{fig:sn-msw} according to
Ref.~\cite{Jegerlehner}. In the case $\tau=1$ oscillations have no
effect so that this is identical to the standard no-oscillation
scenario. Apparently the measured signal characteristics are nearly
incompatible with the theoretical predictions which are indicated by
the hatched rectangle in Figs.~\ref{fig:sn-vacuum} and
\ref{fig:sn-msw}. This effect is due to the surprisingly low
energies of the events in the Kamiokande detector.

According to Eq.~(\ref{eq:energies}) a typical value for the relative
spectral temperature is $\tau=1.7$. According to
Fig.~\ref{fig:sn-vacuum} this would be inconsistent with the vacuum
solution to the solar neutrino problem because the expected event 
energies in the detector would have been even larger than in the
standard case, contrary to the relatively low energies that were
actually observed. Put another way, if the vacuum solution to the
solar neutrino problem is borne out by future experiments, there is a
serious conflict between the SN~1987A observations and theoretical
predictions. 

For the large-angle MSW solution the conflict is less severe
(Fig.~\ref{fig:sn-msw}). For such mixing parameters the flavor
evolution is adiabatic in the supernova envelope so that propagation
eigenstates emerge from the surface which do not oscillate between the
SN and us. However, on the path through the Earth to the detectors,
matter-induced ``regeneration effects'' partly undo the spectral
mixture that emerged from the supernova, i.e. partly restore the
original source spectra, reducing the overall impact of neutrino
oscillations.

In summary, the comparison of the SN~1987A neutrino observations with
theoretical predictions disfavor the large-angle solutions to the
solar neutrino problem, even though the data are too sparse to reach
this conclusion ``beyond reasonable doubt.''


\section{Discussion and Summary}

The minimal picture of neutrinos as espoused by the particle-physics
standard model is still compatible with all established experimental,
astrophysical, and cosmological evidence. Of course, even such minimal
neutrinos would play a dominant role for the dynamics of the early
universe, of supernova explosions, and for the energy loss of evolved
stars. In the absence of any compelling theoretical reason for
neutrinos to be truly massless it is commonplace to assume that they
do carry small masses and that the flavors mix. Cosmology provides by
far the most restrictive limit of about $40\,\rm eV$ on the mass of
all sequential flavors. This limit cannot be circumvented by decays
unless neutrinos interact by new forces which allow for ``fast
invisible'' (i.e.\ nonradiative) decays. Therefore, the assumption of
neutrino masses in excess of about $40\,\rm eV$ is tantamount to the
assumption of a significant extension of the standard model in the
neutrino sector, an extension for which there is no compelling
theoretical motivation. If neutrinos have masses at all, I think it is
a safe bet to assume that their masses obey the cosmological limit.

Neutrinos are unfashionable dark matter candidates because of the
well-known problems of a hot dark matter cosmology if one assumes that
structure forms by gravitational instability from something like a
Harrison-Zeldovich spectrum of initial density perturbations. For the
time being, the standard cold dark matter picture works impressively
well even though it appears to overproduce structure on small
scales. This problem can be patched up by a number of different
modifications, one of them being a hot plus cold dark matter scenario
with a neutrino component corresponding to
$m_{\nu_e}+m_{\nu_\mu}+m_{\nu_\tau}\approx 5\,\rm eV$. However, it
looks unlikely that this sort of scenario can be unambiguously
identified by cosmological methods alone. Even the most ambitious
future cosmic microwave sky maps probably will not be able to identify
this model unambiguously in view of the remaining uncertainty in other
cosmological parameters.

Depending on the exact mixing parameters, neutrino oscillations can
have very severe consequences for supernova physics and the signal
interpretation of SN~1987A or a future galactic supernova. Especially
for neutrino masses in the cosmologically interesting regime,
oscillations may affect the explosion mechanism and r-process
nucleosynthesis in the hot bubble between the neutron star and the
advancing shock wave. However, the current understanding of SN physics
is too uncertain and the SN~1987A data are too sparse to tell if
neutrino oscillations are either required or excluded. Still, it
remains fascinating that a neutrino mass as small as a few eV has {\it
any\/} significant consequences outside of cosmology.

In summary, even though massive neutrinos may play an important role
in cosmology and supernova physics, realistically we will know if this
is indeed the case only by more direct measurements.  With the
possible exception of neutrinoless double $\beta$ decay experiments,
the only fair chance to positively identify neutrino masses is by
oscillation experiments. In principle, oscillations can explain the
atmospheric neutrino anomaly, the LSND $\overline\nu_e$ excess counts,
and especially the solar neutrino problem. However, a simultaneous
explanation of all three phenomena by oscillations is barely possible
and somewhat implausible. It is my personal opinion that the current
round of solar neutrino experiments holds the most realistic promise
of producing uncontestable evidence for neutrino physics beyond the
narrow confines of the standard model.


\end{document}